\documentclass[prl,floats,twocolumn]{revtex4-1}
\usepackage{amsmath}
\usepackage{bbm}
\usepackage{graphicx}

\begin{document}

\title{Electron-induced rippling in graphene}
\author{P. San-Jose$^a$, J. Gonz\'alez$^a$ and F. Guinea$^b$  \\}
\address{$^a$Instituto de Estructura de la Materia,
        Consejo Superior de Investigaciones Cient\'{\i}ficas, Serrano 123,
        28006 Madrid, Spain\\
        $^b$Instituto de Ciencia de Materiales de Madrid,
        Consejo Superior de Investigaciones Cient\'{\i}ficas, Cantoblanco,
        Madrid, Spain}

\date{\today}

\begin{abstract}
We show that the interaction between flexural phonons, when corrected by the exchange of electron-hole excitations,
may place the graphene sheet very close to a quantum critical point characterized by the strong suppression of the bending rigidity of the membrane.
Ripples arise then due to spontaneous symmetry breaking, following a mechanism similar to that responsible for the condensation of the Higgs field in relativistic field theories. In the presence of membrane tensions, ripple condensation may be reinforced or suppressed depending on the sign of the tension, following a zero-temperature buckling transition in which the order parameter is given essentially by the square of the gradient of the flexural phonon field.
\end{abstract}

\maketitle



{\em Introduction.---}
The discovery of graphene \cite{Novoselov:S04,Novoselov:N05,Zhang:N05}, the two-dimensional (2D) metallic crystal made of a carbon monolayer, has spurred a flurry of research activity due to its unique electronic properties and potential applications \cite{Neto:RMP09}. Its mere existence came somewhat as a surprise, since it was widely thought that thermal fluctuations should inevitably destabilize atomic-thin crystals. 
Graphene, however, is not only stable, but surprisingly robust and stiff \cite{Lee:S08}, crucially due to the strong and directional $\sigma$ bonds holding the carbon atoms together. From a theory point of view, these bonds provide the fixed connectivity that protects graphene's long-range crystalline order against thermal dislocations, and the in-plane shear modulus that allows it to remain flat below the crumpling transition temperature\cite{Nelson:JDP87,Doussal:PRL92}.

A remarkable and unexpected property observed in this flat phase, moreover, is its strong tendency to develop \emph{ripples}, or long wavelength modulations of the out-of-plane displacements \cite{Meyer:N07}. X-ray diffraction experiments on suspended graphene revealed that this rippling is not simply a thermal excitation of flexural phonons around a flat average, but a true broken symmetry of the system that \emph{freezes} into a corrugated average configuration \cite{Meyer:N07}. Ripples are expected to have a significant impact on electronic transport in graphene~\cite{Katsnelson:PTRSA08}.

The origin of graphene's rippling has been heavily debated. In exfoliated graphene, ripples are correlated to some extent with the irregularities of the substrate on which the graphene sheet is deposited \cite{Ishigami:NL07}. But there is also evidence that they may arise in part as an effect
intrinsic to the 2D membrane \cite{Meyer:N07}. It has been proposed, for example, that variable length $\sigma$-bonds characteristic of carbon\cite{Fasolino:NM07}, or $sp^3$ orbital rehybridization driven by electron-phonon coupling \cite{Kim:EL08} could underlie ripple formation.

In this work we argue that $\pi$-electrons in graphene play a crucial role in the ripple formation mechanism, since their coupling to flexural modes may lead to a strong renormalization of the elastic parameters.
Above certain value of the  electron-phonon coupling, we find that the system gets very close to a quantum critical point characterized by the strong suppression of the bending rigidity. We propose that, in this situation, ripples arise naturally by spontaneous symmetry breaking, following a mechanism similar to that responsible for the condensation of the Higgs field in relativistic field theories. We are able to characterize in this way a buckling transition that is present at the level of the zero-temperature field theory, with an order parameter controlled by the sign of the tension in the membrane, and leading to a predicted buckled phase consistent with experimental observations.

{\em Softening of flexural phonons.---}
We describe the elastic deformations of a graphene sheet by the vector field $\mathbf{u} = (u_1, u_2, h)$, where $u_1, u_2$ represent the in-plane displacement with respect to the equilibrium position and $h$ is the out-of-plane shift.
The elastic energy of the membrane is expressed in terms of the strain tensor
\begin{equation}
u_{ij}(\mathbf{r}) = \frac{1}{2} (\partial_i u_j + \partial_j u_i +
                       \partial_i h \: \partial_j h  )
\end{equation}
Coordinate $\mathbf{r}$ runs over the 2D graphene membrane.
The bare parameters involved in the continuum elasticity model for graphene are the mass density $\rho $,
the bending rigidity ($\kappa\approx 1$ eV) and the in-plane shear ($\mu\approx 9 \mbox{ eV/\AA}^2$) and bulk ($\mu+\lambda\approx 12.6 \mbox{ eV/\AA}^2$) moduli. The action for the phonon fields reads
\begin{equation}
S_\mathbf{u}= \frac{1}{2} \int dt \: d^2 r (\rho  (\partial_t \mathbf{u} )^2 -  \kappa (\nabla^2 h)^2 - 2\mu u_{ij}^2 - \lambda  (u_{ii})^2  )
\label{s1}
\end{equation}

Respect to other conventional insulating membranes, a novel effect in graphene is that the electronic carriers couple to the displacement of the sheet,
in such a way that it gives rise to another source of
interaction between the phonon fields\cite{Gazit:PRB09,Gonzalez:NJP09}. 
We will consider the
effect of electrons from the $\pi $ bands of graphene, represented
by the field $\Psi (\mathbf{r})$, on the renormalization of elastic parameters.
The strongest coupling between electrons
and phonons comes from the on-site deformation potential \footnote{For simplicity, we consider only the coupling to the trace of the stress tensor. There is another coupling to the electron velocity operator. Apart from numerical factors, our analysis is valid for any of the two contributions to the deformation potential.},
\begin{equation}
S_{\rm e-ph} = \int dt \: d^2 r \; \Psi^{\dagger} (\mathbf{r}) \Psi (\mathbf{r}) \:
( g_{\rm in} \partial_i u_i + g_{\rm out} \partial_i h \partial_i h )
\label{s3}
\end{equation}
with general momentum dependent couplings $g_{\rm in/out}(\mathbf{q})$.
Rotational invariance implies that at small momenta $g_{\rm in}$ and $g_{\rm out}$ must coincide. A microscopic analysis shows however that they can be very different at large momenta, with a general behavior $g_{\rm in/out} (\mathbf{q}) = g(1 - c_{\rm in/out} |{\bf q}|^2 + \ldots )$\cite{Gonzalez:NJP09}. 
Detailed electron-phonon calculations have shown indeed that $g_{\rm in}$  has very small values $\lesssim 1$ eV at the $K$ point of the Brillouin zone\cite{Piscanec:PRL04}.

\begin{figure}[t]
\begin{center}
\includegraphics[width=3.2cm]{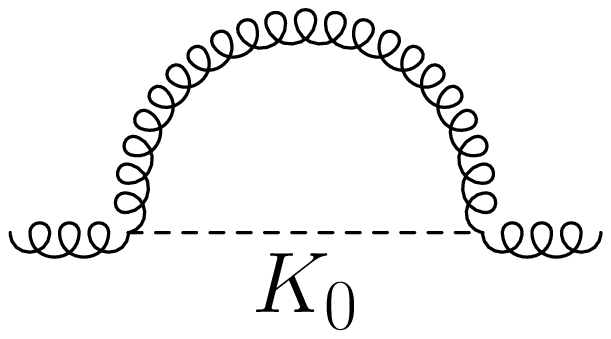}
\hspace{0.8cm}
\includegraphics[width=3.2cm]{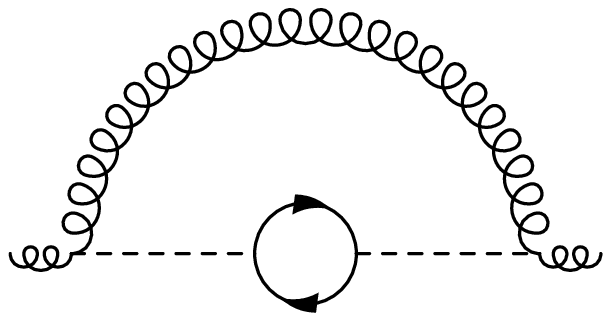}\\\vspace{0.5cm}
 \hspace{0.3cm}  (a) \hspace{3.6cm} (b)
\end{center}
\caption{Lowest order corrections to the self-energy of flexural phonons
(represented by a curly line) arising from (a) the four-phonon interaction
$K_0$ and (b) the exchange of electron-hole excitations (represented by the
bubble with fermion lines).}
\label{one}
\end{figure}

\begin{figure}[t]
\begin{center}
\includegraphics[width=4.2cm]{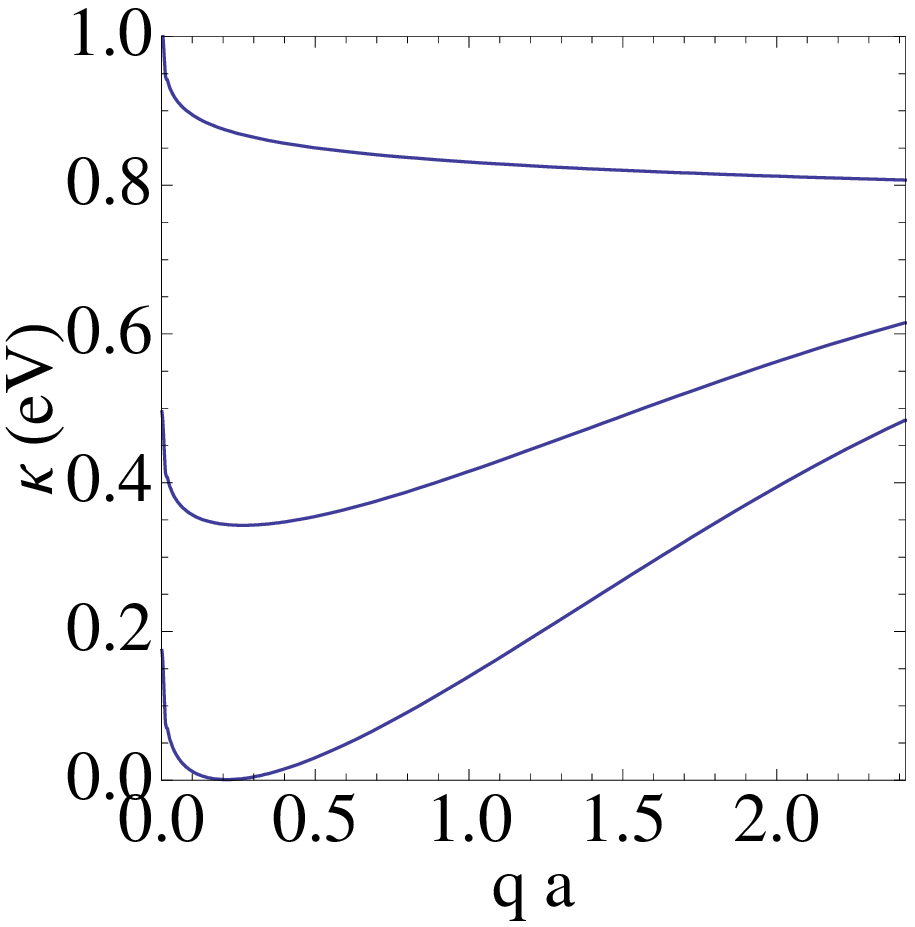}
\includegraphics[width=4.2cm]{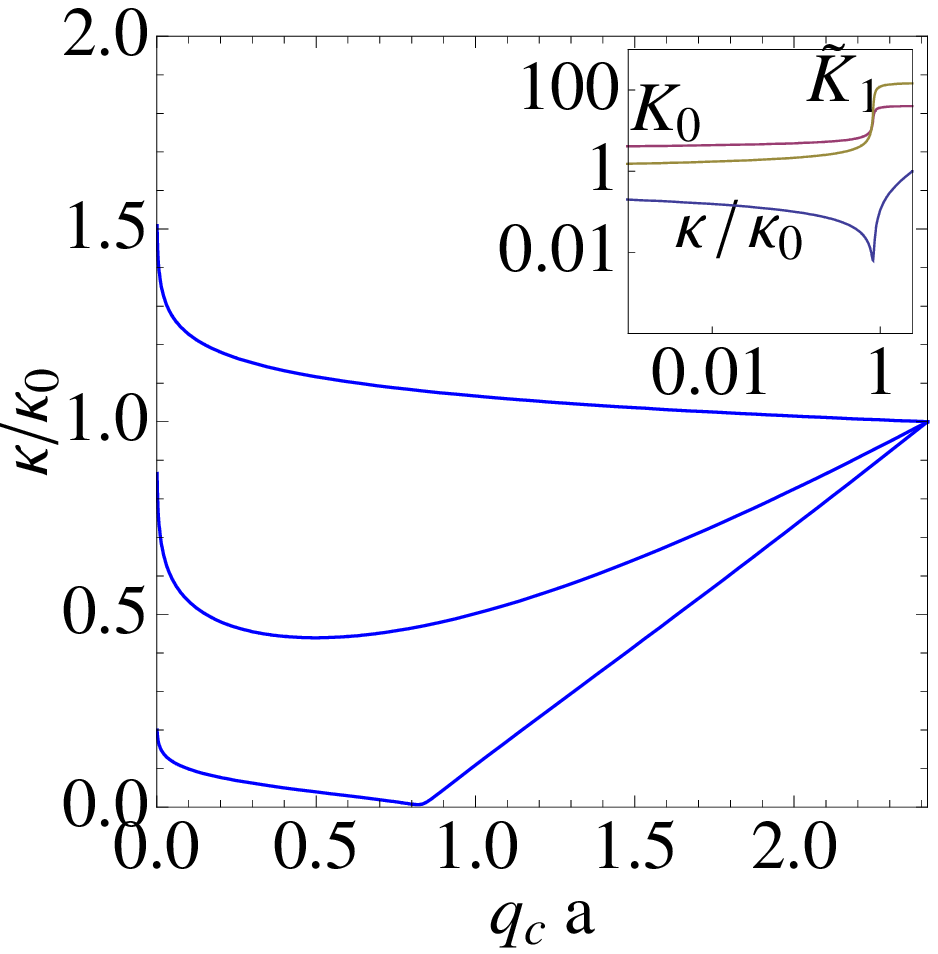}\\
 \hspace{0.3cm}  (a) \hspace{3.6cm} (b)
\end{center}
\caption{(a) Effective momentum dependence of $\kappa $ obtained
from Eq. (\ref{integ}) for $\kappa_0 = 0.8$ eV, $K_0 = 20.5 \: \mathrm{eV/\AA}^2$, and
values of
the deformation potential $g$ equal (from top to bottom)
to 0, 21 and 23 eV.
(b) Scaling
of $\kappa $ in the RG approach, for bare values of $g$ equal (from top to bottom) to 0, 27 and 30 eV (for which the scaling
of $K_0$ and $\widetilde{K}_1 = K_1/q_c$ 
is also displayed in the inset). $a$ is the C-C distance.}
\label{two}
\end{figure}

The energy of the electronic excitations is proportional to the Fermi
velocity $v_F$ and (for a given momentum) much larger than that of other
modes in the problem. Thus, we can first integrate out the electron degrees
of freedom, which leads to a phonon interaction due to the exchange of electron-hole pairs, proportional to the static charge susceptibility $\chi ({\bf q}, 0)\sim -|{\bf q}|/v_F$ \footnote{The analysis of the corrections from the linear dependence on momentum of the classical charge compressibility has been made in Ref. \onlinecite{Gazit:PRB09}. Here we focus instead on the quantum compressibility, starting from elementary electronic excitations.}. One may then integrate out the in-plane phonons, and arrive at the action for the flexural modes
\begin{eqnarray}
S  & = &   \frac{1}{2} \int \frac{d^2 q}{(2\pi)^2} \frac{d\omega}{2\pi} \; (\rho \: \omega^2  -
\kappa {\bf q}^4  ) h({\bf q}, \omega) \: h(-{\bf q}, -\omega)
\nonumber                                                     \\
  &  &  -  \frac{1}{2} \int \frac{d^2 q}{(2\pi)^2} \frac{d\omega}{2\pi}  \;
K(q) \widetilde{u}({\bf q}, \omega) \widetilde{u}({-\bf q}, -\omega)
\label{act}                            
\end{eqnarray}
where $\widetilde{u}({\bf q}, \omega)$ stands for the Fourier transform of
$\frac{1}{2}P_{ij} \partial_i h  \partial_j h$, $P_{ij}$ being the transverse
projector \cite{Nelson:JDP87}, and
\begin{equation}
K(q) = 2\mu + \lambda - g_{\rm out}^2 |{\bf q}|/v_F
 - \frac{(\lambda - g_{\rm in} g_{\rm out}  |{\bf q}|/v_F)^2 }
                               {2\mu + \lambda - g_{\rm in}^2 |{\bf q}|/v_F }
\label{kq}
\end{equation}

An electron-induced bulk instability occurs if the pole in the coupling function $K(q)$ falls inside the first Brillouin zone, leading to a vanishing velocity of in-plane longitudinal phonons. The possibility that such an instability could explain ripple formation\cite{Gazit:PRB09} is ruled out however in a real graphene system, since experimentally the in-plane integrity of rippled graphene is intact. But even if the coupling $g_{\rm in}(\mathbf{q})$ is not strong enough to destabilize the bulk, exchange of particle-hole pairs can make the flexural phonon coupling function $K(q)$ negative (i.e. attractive), which indicates a potential instability of out-of-plane displacements. The effect of the negative terms in Eq. (\ref{kq}) is only
significant at short wavelengths, and vanishes in the limit ${\bf q} \rightarrow 0$. Then, in order to study the competition between positive and negative couplings in Eq. (\ref{kq}),
we may approximate this coupling function by the constant term
$K_0$ and the dominant powers of
$|{\bf q}|$
\begin{equation}
K(q) \approx K_0 - K_1 \: |{\bf q}| - K_2 \: |{\bf q}|^2
\label{linear}
\end{equation}
with $K_1\propto g^2$ and $K_2\propto g^4$.

The resulting interactions tend to enhance or suppress the low-energy rigidity $\kappa $ in opposite directions depending on their repulsive or attractive character. This is encoded into the corrections to the phonon
self-energy $\Sigma ({\bf q}, \omega )$ represented in Fig. \ref{one}.
Computing the full propagator of the flexural phonons from the expression
$D^{-1} ({\bf q}, \omega ) = \omega^2 - (\kappa_0 /\rho) {\bf q}^4
 - \Sigma ({\bf q}, \omega )$, we find that the dressed bending
rigidity $\kappa ({\bf q})$ is given by the self-consistent equation
\begin{equation}
\kappa ({\bf q})  =
\kappa_0  + \frac{1}{8 \pi^2 }  \int d^2 p  \sin^4 (\phi_{q,p})
   \frac{K (|{\bf q}-{\bf p}|)}{|{\bf q}-{\bf p}|^4}
   \frac{|{\bf p}|^2}{\sqrt{\rho \kappa({\bf p}) }}
\label{integ}
\end{equation}

\begin{figure}[t]
\begin{center}
\includegraphics[width=3.2cm]{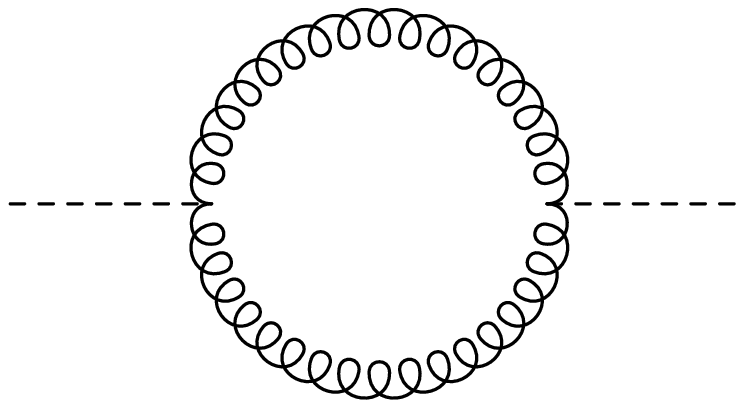}
\hspace{0.8cm}
\includegraphics[width=3.2cm]{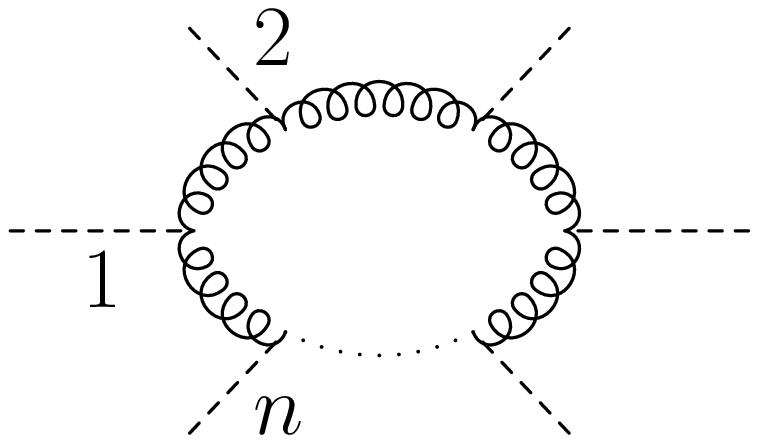}\\\vspace{0.5cm}
 \hspace{0.3cm}  (a) \hspace{3.6cm} (b)
\end{center}
\caption{(a) Exchange of flexural phonons (curly lines) leading to logarithmic
cutoff-dependence of the interactions $K_0, K_1$ and $K_2$. (b) Generic form of
the diagrams building up the power series of the $\sigma $ field (dashed lines) in the effective potential, of which (a) is the first term.}
\label{three}
\end{figure}

This integral equation can be solved
in different regimes of the couplings, see Fig. \ref{two}(a). We observe that, for a
sufficiently large value $g^*$ of the deformation potential,
$\kappa ({\bf q})$ vanishes at a certain momentum, bouncing back
for smaller values of $|{\bf q}|$. 
In this
picture, $g^*$ plays the role of critical coupling above which
we would find negative values of the bending rigidity, suggesting a flexural instability in the system. In this situation, a renormalization group (RG) analysis may be applied in order to also capture the low-energy
scaling of the couplings $K_n$, which is quite
significant when $\kappa $ becomes very small.

The RG scheme proceeds by progressive integration of energy shells, starting from a
high-energy cutoff $E_c = \sqrt{\kappa_{0} /\rho } \: q_c^2$,
to approach the low-energy regime. The bare couplings $K_n$
are then corrected to lowest order by the exchange of
two flexural phonons given by the diagram in Fig. \ref{three}(a), which
shows a logarithmic dependence on the cutoff $q_c$. This can be absorbed into
a renormalization of the effective couplings, leading to the scaling with the
running cutoff $q_c \rightarrow 0$
\begin{equation}
q_c \frac{\partial K_n }{\partial q_c } =  n \: K_n
  + \frac{3}{64 \pi} \sum_{i+j=n} \frac{K_i  K_j}{\sqrt{\rho } \kappa^{3/2} }
\label{grg}
\end{equation}

A scaling equation can be also written for the rigidity $\kappa $, as the
self-energy diagrams in Fig. \ref{one} display a logarithmic dependence
on the cutoff $q_c$,
\begin{equation}
q_c \frac{\partial \kappa }{\partial q_c} =
 - \frac{3}{16 \pi} \frac{K_0}{\sqrt{\rho \kappa} }
 + \frac{3}{16 \pi} \sum_{n=1,2} \frac{K_n}{\sqrt{\rho \kappa} }
\label{kprg}
\end{equation}

The behavior of $\kappa $ obtained from the scaling equations is shown in
Fig. \ref{two}(b) for different bare values of $g$. For small values of this
coupling, we see that the effective rigidity has a steady increasing trend
as $q_c \rightarrow 0$. There is however a regime,
corresponding to $g \sim 20-30$ eV, in
which $\kappa $ is significantly reduced due to the electron-phonon interactions encoded into $K_{1,2}$.
For sufficiently strong $g$, $\kappa $ will be suppressed by
several orders of magnitude, eventually driving the system into a strong-coupling regime $K\gg \sqrt{\rho \kappa ^3}$, and then bouncing back to become trapped in a range of very small values. As in the self-consistent approach, this points to a transition into a new regime in which flexural modes are destabilized. Although the deformation potential is subject to a significant uncertainty in graphene, with $g \sim 4 - 30$ eV \cite{VCKAC01,Suzuura:PRB02,BSHSK08,CJXIF08,HS08}, the bare values for the RG approach should correspond to the large estimates in that range, which apply to the interaction with high-energy electronic states\cite{Suzuura:PRB02}.
We note that the possibility of such an electron-mediated rippling instability is consistent with the observed absence of ripples in fluorinated graphene\cite{Geim:PC}, which differs with respect to clean graphene in the lack of
conduction electrons from the $\pi $ band.

{\em Spontaneous symmetry breaking.---}
Close to vanishing $\kappa$, the renormalized action for $h$ takes the form of a critical scalar field theory with repulsive interaction $K_0$ and a reduced cutoff $E_c$. The ground state properties of this theory are nontrivial due to the role of quantum corrections, which may destabilize the classical (flat) solution leading to a ground state with broken symmetry. The standard technique to compute the equilibrium configuration of the system is the minimization of the effective action
$S_{\rm eff} (h)$ \cite{Coleman:PRD73}. While symmetry breaking has been described before in the statistical mechanics of polymerized membranes\cite{Guitter:PRL88}, here we deal with the computation of the effective action of the quantum theory at zero temperature. 
This can be accomplished by introducing first an auxiliary field $\sigma$ to decouple the four-phonon interaction in (\ref{act}), allowing to express the 
interaction term as
\begin{equation}
S_{\rm int}   =  \frac{1}{2} \int 
         \frac{d^2 q}{(2\pi)^2} \frac{d\omega}{2\pi} \; \sigma ({\bf q}, \omega)  
\left(  \sigma ({-\bf q}, -\omega)
 - 2 \sqrt{K_0}  \widetilde{u}({-\bf q}, -\omega) \right)
\label{s0}
\end{equation}
The effective action is found by decomposing $h$ into an average field 
$h_{\rm av}$ and quantum fluctuations $\widetilde{h}$. Quantum corrections are obtained by integrating 
over $\widetilde{h}$, what can be accomplished exactly in the formal limit 
of a large number of dimensions $d$ of the ambient space containing the membrane.

At large $d$, the effective action is built from diagrams with just one loop of fluctuating 
$\widetilde{h}$ fields, as represented in Fig. \ref{three}(b). In the case of static field 
configurations 
$\widetilde{u}_{\rm av} ({\bf q}, \omega) = \delta (\omega ) \widetilde{u}_{\rm av} ({\bf q})$ 
and 
$\sigma ({\bf q}, \omega) =  \delta(\omega) \sigma ({\bf q})$,
we get the contribution to the effective action
\begin{eqnarray}
\lefteqn{  iS_{\rm eff}^{(1)}   =   \sum_{n=2}^{\infty }   
     \int_{|{\bf q}_i|<\Delta} \prod_{i=1}^n \frac{d^2 q_i }{(2\pi)^2 }
 \:   \sigma ({\bf q}_i) \:  \delta \left( \sum  {\bf q}_i \right)    }
                                                      \nonumber          \\
&  &     \frac{1}{n} \frac{K_0^{n/2}}{(2\pi)^n}   \int_{|{\bf p}|>\Delta}  
    \frac{d^2 p}{(2\pi)^2} \frac{d \omega_p}{2\pi }  \sin^2 (\phi_{p,q_i}) 
           \frac{p^{2n}}{(\rho \omega_p^2 - \kappa p^4 )^{n}}  \;\;\;\;\;
\label{ser}
\end{eqnarray}
We focus here on the field configurations as $\Delta \rightarrow 0$ and 
the momentum of $\sigma ({\bf q})$ goes to zero, since this is the regime where terms with higher powers of
the field become increasingly infrared divergent.

The perturbative series has to be summed first in (\ref{ser}) to obtain a sensible
result in the limit $\Delta \rightarrow 0$. After factoring out the volume of 
the space, we obtain the contribution to the effective potential 
\begin{eqnarray}
iV_{\rm eff}^{(1)}  &  =  & \int \frac{d^2 p}{(2\pi )^2} \frac{d \omega_p}{2\pi }
 \left( - \frac{\sqrt{K_0}}{2\pi}  \sigma_0
  \frac{p^2}{-\rho \omega_p^2 + \kappa p^4}   \right.      \nonumber           \\
   &   &  \left.  
 + \log \left(1 +  \frac{\sqrt{K_0}}{2\pi}  \sigma_0
  \frac{p^2}{-\rho \omega_p^2 + \kappa p^4} \right)   \right)
\label{log}
\end{eqnarray}
where $\sigma_0 \equiv (\Delta /2\pi )^2 \sigma ({\bf q}\rightarrow 0)$.
Performing now the loop integral,
we get from the sum of the zero-loop and the one-loop effective potential
\begin{eqnarray}
V_{\rm eff}  &  =  &   \frac{1}{8\pi^2 } \left( - \sigma_0^2
 + 2 \sqrt{K_0} \sigma_0  \tfrac{\Delta^2}{(2\pi )^2}  
                      \widetilde{u}_{\rm av} ({\bf q} \rightarrow 0) \right) 
                                                                   \nonumber     \\
   &   &   + \frac{1}{8\pi^2 } \frac{K_0}{16\pi \sqrt{\rho} \kappa^{3/2}}  \sigma_0^2
  \left(  
   \log \left(\frac{\sqrt{K_0}}{8\pi \sqrt{\rho \kappa}} \frac{\sigma_0}{E_c} \right) 
                               + \frac{1}{2}      \right)  \;\;\;\;\;\;
\label{eff}
\end{eqnarray}

Quite remarkably,
Eq. (\ref{eff}) exactly reproduces the structure of the effective
potential for a relativistic scalar field theory in 3+1 dimensions\cite{Coleman:PRD73}. 
This means that our model actually follows the same mechanism of symmetry 
breaking characteristic of a Higgs field 
in particle physics. In close analogy with the analysis of Ref. \onlinecite{Coleman:PRD74},
the slightest tension in our model will add to the potential a term 
proportional to $\gamma \widetilde{u}_{\rm av} ({\bf q} \rightarrow 0)$, destabilizing
the minimum at the origin for $\gamma < 0$.
The mode $\sigma_0$ can be integrated by the saddle-point method, showing that 
$V_{\rm eff}$ will get then the typical ``mexican-hat" shape as a 
function of $\widetilde{u}_{\rm av}$, implying a minimum
with $\nabla h_{\rm av} \neq 0$.

We find therefore that there is a buckling transition between
the regimes of positive and negative tension in graphene. The nature of the critical 
(tensionless) theory is less clear, as there is still controversy about 
whether a scalar field theory may undergo symmetry breaking starting with a massless scalar field.
Anyhow, in a typical experimental setup, graphene is held in place by a scaffold or
a substrate, which induces certain amount of tension in the membrane. This has been estimated
to be of the order of $\sim 1 \; {\rm eV/nm}^2$ \cite{Kim:EL08}. 
In our analysis of the symmetry breaking,
a negative tension $\gamma $ shifts the minimum of the effective potential to 
$(\Delta/2\pi )^2\widetilde{u}_{\rm av} ({\bf q}\rightarrow 0) = -\gamma /2K_0 $. This quantity 
turns out to be of the order of $\sim 10^{-4}$ with the above estimate.
Assuming that $(\Delta/2\pi )^2 $ scales in inverse proportion to the volume of the space, 
we obtain that the average values of $|\nabla h_{\rm av}|$ must be of the 
order of $\sim 10^{-2}$, which is consistent with the typical aspect ratio of ripples 
in graphene. 

In the tensionless limit, our model also shares with massless scalar field theories 
in 3+1 dimensions the intriguing feature that the effective potential becomes complex 
beyond a certain value of the field. This kind of instability
has led to speculate about the possibility that the absolute minimum of the effective
potential may be away from the origin in the critical theory\cite{Coleman:PRD74}. 
We hope that the connection
between buckling and the Higgs condensation may
shed light into this question, while carrying out experiments at the low-energy scale 
of graphene.

{\em Conclusion.---}
We have seen that the interaction between flexural phonons mediated by particle-hole excitations may place the graphene sheet very close to a quantum critical point
characterized by the strong suppression of the bending rigidity of the membrane. The same effect can be expected from other degrees of freedom which couple to the lattice deformations, such as absorbed impurities.
The effective potential of the zero-temperature theory displays then
a mechanism of symmetry breaking that parallels that of the Higgs field in particle
physics, the role of order parameter being played in graphene by the
square of the gradient of the flexural phonon field. We have found that the system
must be unstable towards a buckling transition that is the analogue of Higgs condensation,
showing another way to employ graphene as a test ground of fundamental concepts in theoretical physics.

We acknowledge financial support from MICINN (Spain) through grant
FIS2008-00124/FIS.

\end{document}